# The Dynamics of Exchanges and References among Scientific Texts, and the *Autopoiesis* of Discursive Knowledge

Diana Lucio-Arias[1] & Loet Leydesdorff[2]

Amsterdam School of Communications Research (ASCoR), University of Amsterdam
Kloveniersburgwal 48, 1012 CX  Amsterdam, The Netherlands.
[1]Tel.: +31-20- 5253753; Fax: +31-20- 525 3681;  D.P.LucioArias@uva.nl
[2] loet@leydesdorff.net; http://www.leydesdorff.net

**Abstract**
Discursive knowledge emerges as codification in flows of communication. The flows of communication are constrained and enabled by networks of communications as their historical manifestations at each moment of time. New publications modify the existing networks by changing the distributions of attributes and relations in document sets, while the networks are self-referentially updated along trajectories. Codification operates reflexively: the network structures are reconstructed from the perspective of hindsight. Codification along different axes differentiates discursive knowledge into specialties. These intellectual control structures are constructed bottom-up, but feed top-down back upon the production of new knowledge. However, the forward dynamics of diffusion in the development of the communication networks along trajectories differs from the feedback mechanisms of control. Analysis of the development of scientific communication in terms of evolving scientific literatures provides us with a model which makes these evolutionary processes amenable to measurement.

## Introduction

In the field of science studies and its various subfields (e.g., Hackett *et al.*, 2007; Moed *et al.*, 2004), definitions of the units of analysis have molded different approaches through which the sciences can be studied and analyzed. Sociological and anthropological perspectives, for example, have focused on research practices (Latour, 1987; Knorr-Cetina, 1999). Historical reconstructions have based their narratives on the chronology of relevant events in social contexts. From the perspective of the philosophy of science, the main concerns are the epistemological nature and the validity of knowledge claims. Using a model based on scientific literature, information scientists have focused on documents, textual attributes (e.g., author names and references), and their relations.

Relations among authors span a social network, but the relations among documents shape an archive with self-referential dynamics (Burt, 1983; Amsterdamska & Leydesdorff, 1989; Leydesdorff, 1998). These structural dynamics set the conditions for authors to submit new knowledge claims. From this perspective, the submissions provide the variation whereas the evolving networks select deterministically. The specification of the selection mechanisms thus becomes the focus of evolutionary theorizing about the sciences as networked systems.

In the case of the science system, the selection mechanisms operate differently from markets or non-market exchange mechanisms (Nelson & Winter, 1977 and 1982; cf. Whitley, 1984). In this paper, we contribute to the theme of "a science of science" by using Maturana & Varela's (1980) theory of *autopoiesis* or self-organization, and Luhmann's ([1984] 1995) theory of social systems for the specification of how different selection mechanisms are constructed from and feed back upon the process of scientific publishing in recursive loops (Fujigaki, 1998; Maturana, 2000). The "literary model" of science (Price, 1965; Garfield, 1979, at pp. 81f.) enables us to trace publications and their dynamics, and therefore operationalize these evolutionary theories. Both the networks (at each moment of time) and the self-referential loops (over time) can be expected to operate as distributions: uncertainty in these distributions can be measured.

**The *autopoietic* operation of the science system**

Let us assume that the crucial events in scientific communication are publications. Scientific publications formalize communication by relating each submission to the relevant literature. The communication of results obtained in local research labs first serves the further articulation of research questions. Formalization of the communication serves the validation and diffusion of new knowledge claims. Validated knowledge can be used by other scientists as stepping stones for new research and publications. Thus, scholarly activities are embedded in continuous loops of discussing, writing, sharing, and seeking new information (Borgman & Furner, 2002).

Both the communication of scientific findings and their validation by further communication can be considered as crucial to scientific progress. In scientific practices, the two loops can often not be clearly disentangled, but analytically the positioning of a new scientific contribution in a network in terms of (e.g., citing) relations and the recognition of a position from the perspective of hindsight (e.g., being cited) are different. Giddens (1979) called these dynamics of action ("citing") and structure ("being cited") the duality of structure, but failed to specify the selection mechanisms involved (Leydesdorff, 1993).

The two distributions operating as selection mechanisms upon each other are (*i*) the structure in the network at each moment of time, and (*ii*) the development of series of events over time. Variation and selection operate at each moment of time; change and stabilization operate over time. When operating upon each other these two selection environments can be expected to shape historical trajectories in networks of publications. Reflexivity adds (*iii*) a third subdynamic because the meaning of previous communications can be revised by new communications from the perspective of hindsight (Giddens, 1990). Using this third selection mechanism, some previously stabilized trajectories can be selected for meta-stabilization, hyper-stabilization or globalization at a next-order—regime—level (Figure 1).

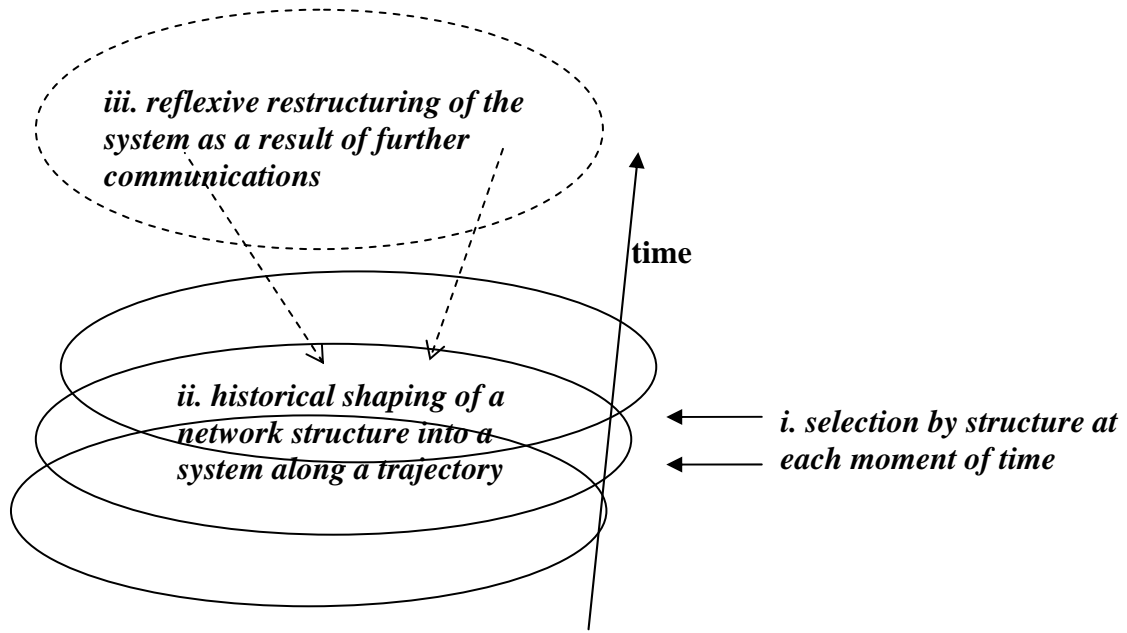

**Figure 1**: Three selection environments operating upon one another.

The interplay between (*i*) variation and selection, (*ii*) retention along trajectories, and reflexivity can be specified by elaborating on Maturana & Varela's (1980) theory of *autopoiesis* or self-organization. While biological systems can provide meaning to information, inter-human communication is more complex. Human languages enable us to *communicate* meaning in addition to information (Luhmann, 1986, 2002a; Leydesdorff, 2000). The communication of knowledge in the sciences—*discursive* knowledge—can be considered as a further refinement of the communication of meaning (Luhmann, 2002b; Leydesdorff, 2007). By using specific codes of communication, scientists are able to develop controlled repertoires (jargons) and process more complexity than by using common (e.g., natural) languages.

Maturana & Varela (1992, at pp. 43f.) defined the autopoietic operation as "the ongoing interactions that produce the components which make up the network of transformations that produces them." In systems theory, *autopoiesis* refers to the ability of an organism or system to reproduce itself: micro-actions creates a network with a next-order structure which feeds back upon the linear flux in a recursive loop (Maturana, 2000). Furthermore, the micro-operations secure the further development of the system by recreating the difference between the system and its environment. Thus, the system can operationally be closed, but the operation provides also a structural coupling to the system's relevant environment(s).

Publications can be considered as the micro-operations of the science system. Their validation using criteria developed by previous publications and their interactions can be considered as a recursive loop. The system of scientific knowledge production and control can thus be considered as a system that accepts and rejects knowledge claims. The scholars involved support this process and condition it by their social and personal characteristics. When the (three) selection processes at the network level reinforce one another, they can result in increased structuration for the agents involved (Giddens, 1984). The recursive loop of codification in the communication can be considered as the (nonlinear) accelerator of science, while the micro-operation of publishing remains essential to science (Stichweh, 1990). Publications interact with one another in terms of exchanges of ideas and arguments and select from previous publications in terms of references. Previous (sets of) publications are continuously recombined and repositioned in the light of new findings in a nonlinear dynamics of communication.

The operation of publishing also reproduces the differentiation between the intellectual system and its social environments: One can expect publications to mean different things locally to a research group or for an individual author's reputation than they do for the intellectual organization of the field. The social and intellectual dimensions co-vary in the events. When publications as micro-actions relate to and build upon one another, a feedback mechanism emerges at the trans-local level of the networks of relations. This resulting structure is dynamic, and its control mechanisms emerge from the system's operations—that is, publications—but as a latent variable—that is, a strategic vector—in the networks of relations. The variation in the knowledge claims is positioned in the literature, and this position can be reflexively refined.

Publications have to be written, linked to the literature, edited, and revised. In these micro-actions, macro-structures resound in an anticipatory mode, and are partially deconstructed and reconstructed, but also to a large extent accepted and reproduced. However, the macro-structures are different from their micro-contexts. While the social contexts of production have hitherto been the main focus of study of social constructivists (Edge, 1979; Hackett *et al*., 2007), the findings are also assigned as *cognitive* assets to a scientific community (Merton, 1973, at pp. 273). The researcher's intention to gain intellectual credit for a novelty feeds back upon the observational reports in the context of discovery (Pinch, 1985). The anticipated allocation of the new findings into a body of knowledge can thus be expected to structure the context of discovery (Myers, 1985).

The three recursive, asynchronous, and interacting selection mechanisms structure next-order developments in the sciences in two directions: the evaluation of scientific contributions integrates knowledge claims into scientific literature (Wouters, 1999); and they reinforce specific codes of communication, including repertoires and references as symbol systems (Small, 1978). The codifications enable scholars to communicate complexity more than in natural languages (Coser, 1975).

The analytical distinction between the intellectual and social organization of scientific knowledge was proposed in the philosophy of science by Karl Popper. Popper ([1935], 1959) distinguished between the context of discovery and the context of justification (Figure 2). While the context of discovery is defined in terms of the social processes that focus upon the creation of research findings, the context of justification is relevant for the processes whereby research findings are transformed into accredited knowledge (Gilbert, 1976). Research findings are validated upon their approval by a research community. This change in the epistemological status of a knowledge claim to accepted knowledge requires mediation between the two contexts in the form of text (Leydesdorff, 2007). Gilbert & Mulkay (1984, at p. 69f.) noted that after the validation of a knowledge claim an additional—rational—repertoire becomes available for legitimizing scientific findings.

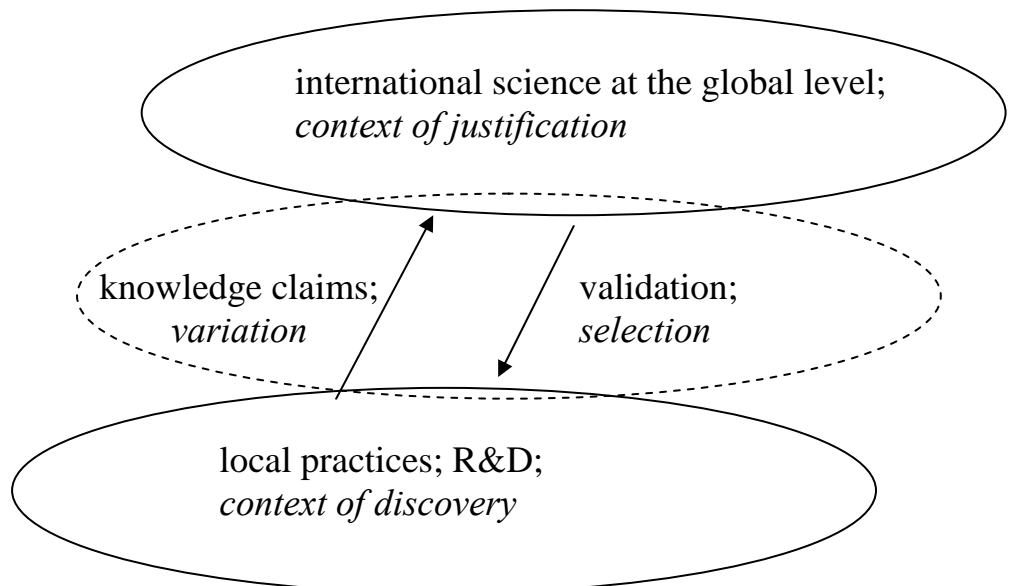

**Figure 2.** Mediation between the context of discovery and the context of justification

Our description of the three selection environments suggests that publications mediate between the two contexts: the constant stream of publications creates and reproduces the intellectual structures that frame new scientific texts (Cozzens, 1985; White *et al.*, 2004). New knowledge claims—the results of scientific practices and research activities—are formulated in texts that are validated through the review process and eventually accepted for publication (or not). If published the new knowledge claims can be integrated into bodies of scientific knowledge and eventually become part of a structuring global repertoire of scientific knowledge. In Giddens's "structuration theory," the resulting structure was conceptualized as shaped over time by memory traces, but structure would only be reproduced in time and space by reflexive recombinations of sets of rules and resources in action (Giddens, 1979). However, reflexivity in communications provides the communications with another selection mechanism.

Reflexivity in human interactions generates a "double hermeneutics" since one can both be enrolled as a participant and/or reflexively understand the configuration (Giddens, 1976). This "double hermeneutics" between a socially-constructed cognitive structure and communicative action can be studied empirically using a model focused upon publishing as the basic operation in producing scientific literature. The validation of a publication changes the epistemological status of the message, and therewith its analytical position in the design.

While the publication is only an element of variation in the network, validation makes the content of the communication part of a reflexively-selecting structure. This next-order layer coevolves with the layer that provides the variation, as its structure. In our opinion, science cannot only be considered as a belief system based in communities of practice and agency (Barnes & Edge, 1982; Bloor, 1976), but also as a system of rationalized expectations contained in scientific literatures. The validation of a new knowledge claim is constrained by what has already gained the status of valid scientific knowledge and thus constitutes the *ex ante* context of justification. Publications make the system operate so that the *ex ante* structures of science are transformed into an *ex post* structure of partially rewritten expectations.

**The communicative turn**

Both Thomas Kuhn and Derek de Solla Price sought to operationalize the scientific enterprise in terms of publications. Kuhn (1962, at pp. 170 ff.) focused on textbooks as constitutive of paradigms, while Price (e.g., 1970) was interested in the potentially discipline-specific dynamics of articles, reviews, and the generation of textbook knowledge. Eventually, Kuhn analyzed the growth of the scientific enterprise in terms of historical discoveries, authors, and the development of paradigms as belief structures, whereas Price took the decisive step of operationalizing scientific developments in terms of the dynamics among texts, their distributions, and the resulting dynamics in networks of scientific communication (Price, 1965).

While publications remain the results of human action, the resulting texts can also be considered as units of analysis in a next-order dynamics. In the first-order network, the authors are the nodes and the texts—or, more precisely, textual attributes—provide the links. Because texts are relational, they can also function as units of operation (Bhaskar, 1975; 1998 at p. 207). In a second-order design, the texts are the units of operation which can be expected to develop codes of communication as latent variables. Discursive knowledge results from these transformative interactions among texts. While texts have to be written and read by people in a first-order dynamics, scientific texts are also stored and circulated beyond the control of specific authors or readers. In this second-order dynamics, the originally communicating scholars can be replaced by (potentially anonymous) peers, while the contents remain in the texts.

The texts contain relevant (and sufficient) information about the authors, such as author names and institutional addresses. Using these representations, it is possible to study the social system of scientific knowledge production and control from the perspective of the textual one. The intellectual feedback from the context of justification structures the stream of scientific publications. For example, the substantive delineations among domains in textual sets are reproduced in terms of their intellectual organization. The reproduction of these delineations from year to year indicates that the intellectual dimension operates with an analytically independent dynamic that guides the organization of publications in journals.

Newly validated scientific claims are made available to the scientific community in publications. Scholars can then use such previous results to support their own research processes. The further articulation of previously validated knowledge claims to support new findings is common practice in research. For example, as Sumio Iijima (2005), the discoverer of nanotubes in 1991, recalled, "when I heard about the discovery of $C_{60}$ in 1985, I thought to myself, SO THAT was the onion-like structure that I saw…" The constant circulation of new knowledge claims and reconstructions stirs the system. Funding, access to datasets, e-mails, and informal communications influence the system of knowledge production as enabling and constraining conditions at each moment of time, but publishing itself can be considered as the autopoietic operation which warrants the science system's integrity and develops its progressive momentum (Fujigaki, 1998).

The structuring contexts of justification and constructive actions at the research level may co-evolve and lead to a relative closure in the communication. Closure leads to specialization and the development of repertoires which enable the scientists involved to communicate specific complexity. The reproduction of structure in the intermediating layer is the result of developments above and below this level, but the textual layer itself can also be expected to develop its own dynamics. For example, journals have to be viable on the market in terms of subscriptions and other earnings.

The intellectual organization cannot be measured directly in the texts like the social addresses of authors and institutions. However, the mutual dependencies among the three dimensions enable us reflexively to reconstruct the dynamics in the intellectual organization. For example, one can consider next-order structures as clusters and components of the networks under study using multivariate statistics. The structural components are not dependent variables, as they would be from a bottom-up and action-oriented perspective, but can be considered as latent constructs (e.g., eigenvectors) which operate top-down when invoked in the observable instantiations. This latent structuring can be measured, for example, by using factor analysis.

Discursive knowledge is generated by exchanges of codified elements among texts (e.g., specific arguments and references) in a next-order dynamics. Socio-cognitive regimes emerge at the supra-individual level from (*i*) streams of publications, (*ii*) their reflexive decompositions and reconstructions in discursive exchanges, and (*iii*) the consequent dynamics in their positions in the network of communications. Although largely beyond control for individual agents, the intellectual structures and dynamics can reflexively be accessed by both participants and analysts. However, the reconstructions at the level of agents cannot be unambiguous since the intellectual organization of the sciences remains in the second-order domain of the construction and consolidation of expectations (Husserl, 1929; Luhmann, 2002b; Leydesdorff, 2007).

In summary, the literary model enables us to study the sciences from a systems perspective. However, the system is nothing more than the operation of different yet specific selection environments on the variation. The selection environments are not externally "given," but can endogenously be generated as cultural constructs by the micro-operation because publication is both a formalized and codified form of inter-human communication. The literary model thus provides us with a functional simplification for studying the self-organization of scientific knowledge in terms of indicators.

By using the literary model for developing indicators, one obtains fruitful heuristics. The focus is no longer on historical cases and single events, but on distributions in sampled document sets. The distributions enable us to test observations against hypotheses. We proceed in the following sections by elaborating on (a) how the basic *mechanisms of growth and change* are generated by the continuous streams of publications; (b) how intellectual structures emerge as *specialties and self-organize*; (c) how this self-organization can be operationalized to measure changes in systems of scientific communication in terms *probabilistic entropy*; and (d) how the science system *interacts* with other social systems.

**a. Mechanisms of growth and change**

Scientists take from and contribute to the commonwealth of scientific knowledge. At the individual level, the importance of making a contribution serves two purposes. The first is to gain recognition for the ownership of a contribution. The intellectual property of a scientist's contribution is acknowledged when it becomes part of the public domain of science (Merton, 1979). Additional value is accorded to a scientific contribution after its publication when other scientific contributions make reference to it or otherwise articulate its knowledge content in the support of new findings. Contributions participate in the development of a discourse independently of the intentions of their authors (Leydesdorff & Amsterdamska, 1990).

Scientific contributions can further be validated when consensus emerges among peers that the contribution is not only original but also relevant to the further development of scientific knowledge. New scientific contributions provide variation (new information) at the emergent and transient grouping of concepts that constitutes the research front (Chen, 2006). By being accepted for publication, the contribution is stabilized as part of the archive. Publications can further be selected and then become part of the codified archive of knowledge. This generalization of what was previously stabilized in a specific domain makes it possible for practicing scientists to use the concepts increasingly without reference to the original publication. The concept then gradually obtains the status of textbook-knowledge, and the specific references can be obliterated (Garfield, 1975; Merton, 1968, at pp. 35 ff.). For example, "oxygen" has a meaning codified in everyday language to the extent that one no longer needs to provide a reference to "(Priestley, 1774)" when using this concept. The previous phlogiston theory was overwritten by the newly constructed concept of "oxygen."

The recursive selection of significant knowledge claims, although potentially latent to the individual scientists involved, upgrades the system of knowledge production in an evolutionary competition. Unlike variation in the communication of scientific findings, the selection of validated knowledge claims is determined by standards emerging from what can be considered as valid knowledge, but at a next-order level. This next-order selection mechanism (articulated perhaps as criteria) is constructed by and feeds back on the publication of scientific texts. Although built bottom-up from the historical process of publishing scientific results, the next-order mechanisms (theories, paradigms) exercise top-down control over the publishing process. The publishing process selects for stabilization in a theoretical context, whereas knowledge contents can be further selected for incorporation at the level of a paradigm. The paradigms provide and structure global "horizons of meaning" (Husserl, 1929; Luhmann, 2002b). Note the plural in "horizons"-- more than a single paradigm is possible. Scientists are additionally able to change and translate among repertoires (Mulkay *et al*., 1983; Gilbert & Mulkay, 1984).

The science system can be considered as a system of operations. New operations provide it with continuity by reproducing existing structures along trajectories. Because of this operational character, the system can change its structure due to operations at the bottom. Historical stabilizations of patterns at the aggregated level serve evolutionary processes of change at the global level (Luhmann, 1995). Evolutionary changes can be incremental or radical: spontaneous breakthroughs can cause the replacement of old paradigms with new ones, henceforth constraining the scientists involved in a different set of contexts. Each selective layer builds upon lower-order ones, but instabilities at the bottom may shake the firmaments spanned by the communication at the top-level of paradigms.

**(a) bottom-up diffusion of the variation**

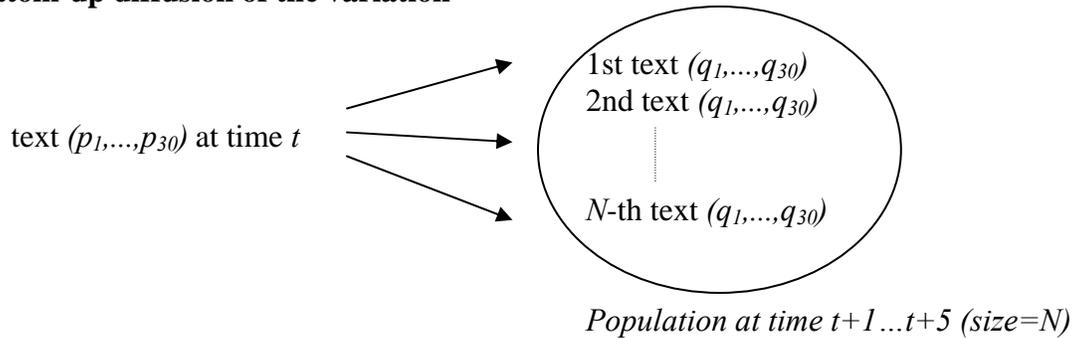

**(b) top-down codification against the axis of time**

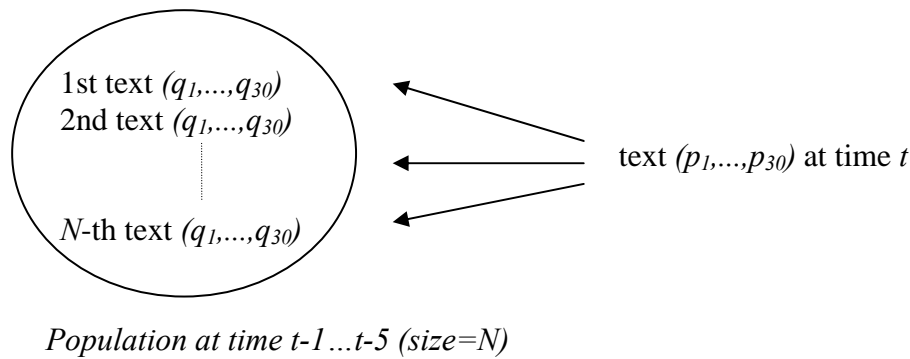

**Figure 3.** Schematic representation of the analysis of diffusion versus codification (Frenken & Leydesdorff, 2000, at p. 336)

Two different dynamics result from the interactions among the selection environments: top-down codification and bottom-up diffusion (Figure 3). Variation provided by new knowledge claims is diffused over time. From the perspective of hindsight and on top of the process of publishing, the reflexive selection of knowledge already incorporated in the scientific literature and anchored in terms of references, provides meaning to the texts. Codification, operating through the selection of already stabilized knowledge claims, preserves the memory of the system by reproducing the existing knowledge structure. Codification can be operationalized in the literary model in terms of references and citations, and the development of specific jargons (e.g., co-occurrences of words and references) in specialties.

**b. Specialization and self-organization**

Mutual shaping in co-evolutions between each two of the three selection mechanisms reinforces the specificity of selections. For example, connections between traditionally unconnected clusters of publications can result in a new specialty (Chen, this issue). However, the axis for codification in the communication is not pre-determined in a system of reflexive communications, and the reflected structures remain uncertain expectations. Using this degree of freedom, differentiation among the codes of communication allows for processing more complexity (Simon, 1973). The differentiated sub-systems can be expected to self-organize their further developments, increasingly using their own codes and criteria for the selection process as they further develop.

Networks of scientific publications exhibit a structure in which disciplinary differences are emergent properties of the networks. When these properties can be reproduced over time, they can also be considered as the latent dimensions of intellectual organization. Change is carried by both the diffusion of literature along trajectories, and the differentiation in the intellectual dimension. However, one can expect the two selection processes—the stabilizing one and the globalizing one—to occur concurrently. To the extent that their interactions are harmonic, scientific discourse can be expected to proceed in terms of proliferating new communications (Smolensky, 1986).

As systems gain in complexity, increasing codification along different axes tends to result in internal differentiation. This makes the system more "viable" because more variety can be processed and controlled (Ashby, 1958; Beer, 1984). Nowadays, "oxygen" has become a relevant subject in a number of specialties. For example, oxygen can be studied as an atom, as a molecule, or in terms of its metabolic functions. Retention mechanisms in the fluxes of communication provide the different discourses with memory of contexts. Previous communications, for example, are stored in libraries and archives. In a diversifying system of communicative interactions, this memory of the system can be expected to remain distributed and therefore in transition. Control is exercised by symbolically generalized criteria that can be made explicit as necessary.

This process of specialization can be considered as a consequence of scientific communication itself. The subsystems do not remain hierarchically nested—although they originate historically from predecessors—but evolve in orthogonal directions as they gain in autonomy because of their increasingly autopoietic operations in terms of specific codes and criteria. Differentiation is the result of the continuous codification of new communication by which previously existing structures can be rewritten into various depths as new knowledge is developed. Whereas participants, policy makers, and research managers can only influence the (institutional and human) conditions of the substantive communications (Spiegel-Rösing, 1973), discursive knowledge results from the interactive dynamics of groupings of texts at the network level. As in the social network, the latent dimensions of networks of communication can be expected to structure the manifest relations. At each moment, the relations between observable variables (vectors) and the latent dimensions (eigenvectors) can be analyzed in terms of network statistics. Over time, one can use probabilistic entropy for the decomposition (Theil, 1972; Leydesdorff, 1995).

The networks provide only the retention mechanisms of a more complex dynamic which evolves from the flows of communication in the orthogonal dimension of intellectual organization. This intellectual organization self-organizes as systems of codified expectations and this emerging structure can be reconstructed reflexively. The reflexive specifications are needed for explaining groupings along intellectual lines in the observable phenomena. For example, one can raise the question why journals can be clustered in maps of science. Intellectual self-organization operates as an independent variable and structures the variation in aggregated networks of journal-journal citations. This socio-cognitive structuration can be operationalized in terms of the eigenvectors of the matrix. Analytically, one is able to distinguish between the dynamics of vectors and eigenvectors, and thus to enrich the explanation.

**c. Probabilistic entropy**

The second law of thermodynamics states that the entropy of an isolated system which is not in equilibrium will tend to increase over time, approaching a maximum value at equilibrium. However, entropy is reduced when the *structuring* of the variation adds to the redundancy. Differentiation and self-organization structure the variation into subsets. A structured system can process more uncertainty; the maximum entropy of the undifferentiated system can be multiplied with the cognitive categories as another dimension of the probability distributions (Brooks & Wiley, 1986; Leydesdorff, 1997). The proliferation of intellectual categories because of the differentiation among the sciences thus expands the maximum entropy of the system.

When the cognitive, textual, and social layers are conceptualized as horizontally super-imposed layers in a hierarchy like in Figure 2, the complexity cannot yet be unfolded; one needs a model of mutually interdependent and mutually shaping dimensions in a "triple helix" of cognitive, social, and discursive subdynamics.

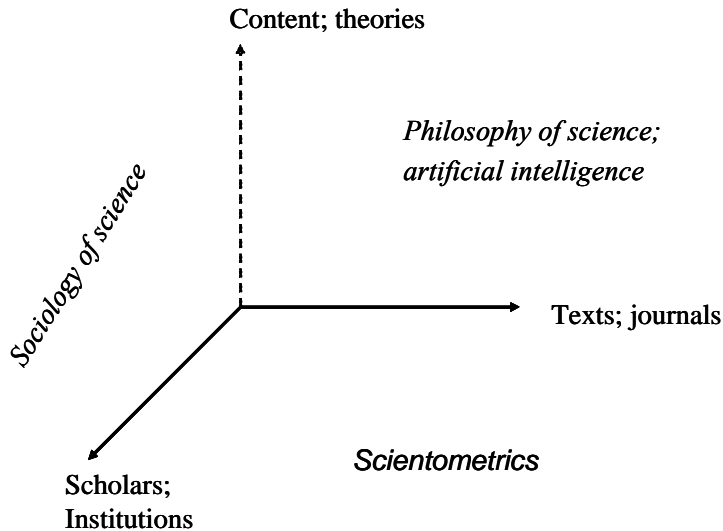

**Figure 4.** Three interacting dimensions of the scientific enterprise. (Source: Leydesdorff, 1995, at p. 3.)

Figure 4 provides a static representation of this triple-helix model. The three layers in Figure 2 are now unfolded as three dimensions. Dynamically, theories, scholars, and texts are mutually dependent; the cognitive, textual, and social dimensions involved in the co-production of scientific knowledge can be expected to interact as spirals. The recursive development in each dimension is reflected in each of the other two subdynamics. Scholars and institutions in the social dimension, for example, recombine texts and cognitions with other social resources such as funding, instruments, and research materials. Authors and institutions are represented in texts which select from theoretical contexts. Similarly, the networks of authors and texts are reflected in an intellectual organization that remains emergent as a structure of expectations.

Despite its intangible character, the intellectual reflection can be expected to feed back in a next round on the previous configuration by enabling and constraining new configurations. The retention of this intellectual structure in the networks can be measured on the basis of the assumption that categories in the cognitive dimension operate within the system and thus leave traces. The evolving structures are reflexively accessible while they are instantiated in the historically observable configurations albeit to variable extents. The networks function as the retention mechanism for the further development in the intellectual dimension.

The trade-off between evolutionary self-organization among the fluxes of communication through the networks and the historical retention and organization in the networks remains an empirical question. The reflexive operation in the cognitive dimension is not directly accessible to measurement, but it can be inferred because it co-constructs the configuration. A configuration is more than the sum of its parts, and this surplus value can be measured as "configurational information" (Abramson, 1963; McGill, 1954; Jakulin & Bratko, 2004; Leydesdorff, 2008; Ulanowicz, 1997; Yeung, 2008). The measure of this configurational information $\mu^*$ (Yeung, 2008, at pp. 51 ff.) captures the possibilities of progression and/or stagnation in the intellectual dimension as a plus or minus sign of the resulting entropy:

$$\mu^*_{xyz} = H_x + H_y + H_z - H_{xy} - H_{xz} - H_{yz} + H_{xyz} \tag{1}$$

Each of the terms in this formula represents a (Shannon) entropy: $H_x = -\sum_x p_x \log_2 p_x$, $H_{xy} = -\sum_{xy} p_{xy} \log_2 p_{xy}$, etc, where $p_x$ represents the probability distribution of attribute $x$ and $p_{xy}$ the probability distribution of attributes $x$ and $y$ combined.[1] The resulting value of the information measure $\mu^*$ can be positive or negative depending on the relative weights of the interactions among the various uncertainties.

Negative configurational information indicates a reduction of the uncertainty prevailing at the level of the configuration or, in other words, structuration (Giddens, 1979). This reduction of uncertainty cannot be attributed to one of the composing subdynamics, but is a result of their interactions which shape the configuration. One can also consider this as a synergy at the systems level, which cannot be localized within the system because it is a result of a configuration. The configuration at each moment of time is a balanced outcome of the trade-off between the differentiated fluxes of communication and retention of this intellectual feedback in networks of communication.

---

[1] In the case of publications, these could be the distribution of title words, cited references, author names, institutional addresses, etc.

By measuring the configurational information as a potential source of negative entropy, one is able to show how the—otherwise intangible—intellectual organization of a document set structures its textual dimension. Note that configurational information provides one method among other possible ones. For example, Sun & Negishi (2008) showed in a recent communication that similar results could be obtained by using partial correlation analysis among three sources of variance (McGill & Quastler, 1955).

**d. Interactions with other social systems**

The communication of scientific results in publications is a process of discrete events. However, the eigenvectors in the networks of relations and, analogously, the eigenfrequencies in the relations among events provide continuous coordinates that shape a hypothesized system. Through this continuous operation of shaping and reshaping, a system of scientific knowledge production emerges which remains dependent for its continuation on the very operations that generate the system. Therefore, the boundaries of the system are also based on the micro-operations. Since these micro-operations are distributed in a system of micro-operations, the boundaries of the system are uncertain. A model based on scientific literature as a composite operation of publishing enables us to specify uncertainty in how the science system *autopoietically* shapes its structure, boundaries, and possible openings for couplings with other networks such as national systems of innovation, markets, etc.

Differentiated communication systems serve the reproduction of complexity in a pluriform society because they allow for a variety of social functions to operate concurrently and in interaction with one another (Luhmann, 1995). However, scientific discourse is only one of the modern coordination mechanisms; economic wealth generation is another. The operation of the system of science results specifically in the production and validation of scientific knowledge and novelty. Knowledge-based innovations are based on interactions between this system of scientific knowledge production and the market as another coordination mechanism.

The boundaries of a self-organizing (sub-)system are produced and reproduced by specifically coded operations. A patent, for example, can be considered as a valuable asset from an economic perspective, while it is an output of the knowledge production system. Although operationally closed in terms of their respective selection criteria, the interactions provide windows of opportunity with specific constraints and benefits. The couplings among scientific knowledge production, economic exchange mechanisms, and political (e.g., national) control provide rich domains for empirical research (Etzkowitz & Leydesdorff, 2000; Leydesdorff & Sun, 2009).

**Conclusions and discussion**

Popper's ([1935] 1959) philosophical distinction between a context of discovery and a context of justification did not appreciate that these contexts are mediated by texts in a process of validation with its own dynamics. In his study entitled *The Intellectual and Social Organization of the Sciences*, Whitley (1984) operationalized the relations between the intellectual and social dimensions in organization-theoretical terms when he characterized the sciences as systems of organized knowledge production and control. In our opinion, knowledge production in the context of discovery indeed can be considered as organized. Self-organization in the intellectual dimension, however, stands orthogonal to the institutional conditions: intellectual organization emerges within the reflexive layer of scientific communications and feeds back in a next-order dynamic.

The scientific communications can be considered as organized in a second-order dynamic on top of a first-order dynamic among historical agents and their organizations. The texts can be attributed to authors and organizations, but the codes of communications can only be attributed to texts. The emerging top-down arrow of control can be expected to constrain the bottom-up arrow of the construction. This feedback is reinforced by the reflexivity in the communicative operation. In the self-organization model the intellectual control centers are no longer organized; one cannot expect a single center of control among the flows of communication. Thus, differentiation can be expected to prevail more than integration.

Three selection mechanisms were specified as operating in this process upon one another: (*i*) the instantaneous positioning of a publication in the network in terms of its (e.g., citation) relations with other texts, (*ii*) the positioning of this publication in relation to other publications along the time axis, and (*iii*) the expected re-positioning and rewriting of the knowledge content of the paper in the further process of the developing discursive knowledge. The three selections correspond with the three contexts: the first selection is performed by the author when constructing the paper as an output in the context of discovery, the second by textual processes such as peer-review organized at the journal level, and the third incorporates the intellectual feedback into the non-linear dynamics of this complex process. Thus, other social contexts are invoked to perform these selections.[2]

From an information-theoretical perspective, the focus is not on the single case, but on the changes in the distributions. The three contexts can be operationalized as uncertainty contained in the distributions in different dimensions. Thus, the selective structures are not reified (Giddens, 1979), but can be specified as uncertainties (Leydesdorff, 1993, 1995). Structures at each moment of time can be operationalized as the dimensions or eigenvectors of the matrices which represent the graph-theoretical networks. The eigenvectors provide the coordinates of a space in which the variation is spanned at each moment of time. The historical dimension adds the notion of a trajectory in a three-dimensional array. The reflexive operation thirdly enables this system to reconstruct itself from the perspective of hindsight, for example, by providing references in new publications to previous literature.

Publications can function as the micro-operators of this self-organization because they contain the elements of these three selection environments: each publication is (*i*) positioned in a codified network of communications because relations are formalized; (*ii*) follows up on previous publications along trajectories; and (*iii*) redefines the intellectual context from an original perspective. By being linked up to the literature, the micro-operation takes part in distributions of micro-operations in the three relevant dimensions: in relation to other publications, previous publications, and in terms of participation in the reflexive reconstructions. The distributions operate in terms of uncertainty which can be measured in terms of probabilistic entropy. The mutual information among three dimensions then provides us with a measure for the self-organization when negative entropy is endogenously produced.

---

[2] Maturana, Varela, and Luhmann use the term "structural coupling" for this coupling to the relevant environments.

By considering publishing as the autopoietic operation of the science system, one can operationalize the dynamics between the context of justification and the context of discovery as a measurable process. From this perspective, the publications and their recursive interactions also shape the boundaries of the operational system. Both analysts and participating scholars can only reconstruct these constructs reflexively because they contain uncertainty. Practicing scientists, for example, have to construct (perhaps intuitively) a model of their specialty when making an assessment of where and how to publish a new knowledge claim.

A model based on scientific literature enables us to specify these emergent systems, their structures, their boundaries, and their couplings with other socially relevant networks (e.g., markets) in terms of uncertainties in relevant distributions. The literary model thus provides us with an empirical turn to address questions that were previously debated in terms of the philosophy of science and/or the sociology of scientific knowledge. In both latter traditions, texts were mainly analyzed qualitatively, as attributes of either scholars or cognitions. Our focus on the discursive dimension as the driver of scientific progress, however, allows us to analyze both people and cognitions in their contexts.

**References**
Abramson, N. (1963). *Information Theory and Coding*. New York, etc.: McGraw-Hill.
Amsterdamska, O., & Leydesdorff, L. (1989). Citations: Indicators of Significance? *Scientometrics 15*, 449-471.
Ashby, W. R. (1958). Requisite variety and its implications for the control of complex systems. *Cybernetica,* 1(2), 1-17.
Barnes, B., & Edge, D. (Eds.). (1982). *Science in Context*. Cambridge, MA: MIT Press.
Beer, S. (1984). The Viable System Model: Its Provenance, Development, Methodology and Pathology. *Journal of the Operational Research Society,* 35(1), 7-25.
Bhaskar, R. (1975). *A Realist Theory of Science*. Leeds: Leeds Books Ltd.
Bhaskar, R. (1998). Societies. In M. Archer, R. Bhaskar, A. Collier, T. Lawson & A. Norrie (Eds.), *Critical Realism: Essential Readings* (pp. 206-257). London/New York: Routledge.
Bloor, D. (1976). *Knowledge and Social Imagery* London, etc.: Routledge & Kegan Paul).
Borgman, C.L., & Furner, J. (2002). Scholarly Communication and Bibliometrics. In B. Cronin (Ed.), *Annual Review of Information Science and Technology*, 36. (pp. 3-72). Medford, NJ: Information Today.
Brooks, D. R., & Wiley, E. O. (1986). *Evolution as Entropy*. Chicago/London: University of Chicago Press.
Burt, R. S. (1983). Network Data from Archival Records. In R. S. Burt & M. J. Minor (Eds.), *Applied Network Analysis: A methodological introduction* (pp. 158-174). Beverly Hills, etc.: Sage.
Chen, C. (2006). CiteSpace: Detecting and visualizing emerging trends and transient patterns in scientific literature. *Journal of the American Society for Information Science and Technology*, 57(3), 359-377.
Chen, C. (2009). Towards an Explanatory and Computational Theory of Scientific Discovery, *Journal of Informetrics* (this issue).

Coser, R. L. (1975). The complexity of roles as a seedbed of individual autonomy. In L. A. Coser (Ed.), *The idea of social structure. Papers in honor of Robert K. Merton* (pp. 237-264). New York/Chicago: Harcourt Brace Jovanovich.
Cozzens, S. (1985). Comparing the sciences: Citation context analysis of papers from neuropharmacology and the sociology of science. *Social Studies of Science*, 15, 127-53.
Edge, D. (1979). Quantitative Measures of Communication in Science: A Critical Review. *History of Science*, 17, 102-134.
Etzkowitz, H., & Leydesdorff, L. (2000). The Dynamics of Innovation: From National Systems and 'Mode 2' to a Triple Helix of University-Industry-Government Relations. *Research Policy,* 29(2), 109-123.
Frenken, K., & Leydesdorff, L. (2000). Scaling Trajectories in Civil Aircraft (1913-1970). *Research Policy,* 29(3), 331-348.
Fujigaki, Y. (1998). Filling the Gap between the Discussion on Science and Scientists' Everyday Activities: Applying the *Autopoiesis* System Theory to Scientific Knowledge, *Social Science Information* 37(1), 5-22.
Garfield, E. (1975). The "obliteration phenomenon" in science—and the advantage of being obliterated. *Current Contents,* December 22, #51/52, 396–398.
Garfield, E. (1979). *Citation Indexing: Its Theory and Application in Science, Technology, and Humanities*. New York: John Wiley.
Giddens, A. (1976). *New Rules of Sociological Method*. London: Hutchinson.
Giddens, A. (1979). *Central Problems in Social Theory*. London, etc.: Macmillan.
Giddens, A. (1984). *The Constitution of Society*. Cambridge: Polity Press.
Giddens, A. (1990). *The Consequences of Modernity.* Cambridge: Polity Press.
Gilbert, G. N. (1976). The Transformation of Research Findings into Scientific Knowledge. *Social Studies of Science*, 6, 281-306.
Gilbert, G. N., & Mulkay, M. J. (1984). *Opening Pandora's Box. A Sociological Analysis of Scientists' Discourse*. Cambridge: Cambridge University Press.
Hackett, E., Amsterdamska, O., Lynch, M., & Wajcman, J. (2007). *New Handbook of Science, Technology, and Society*. Cambridge, MA: MIT Press.
Husserl, E. (1929). *Cartesianische Meditationen und Pariser Vorträge [Cartesian meditations and the Paris lectures]*. The Hague: Martinus Nijhoff, 1973.
Iijima, S (2005). The discovery of carbon nanotubes - Guided by serendipity. Available at http://www.nec.co.jp/rd/Eng/innovative/E1/01.html. (Retrieved on November 30, 2008.)
Jakulin, A., & Bratko, I. (2004). Quantifying and Visualizing Attribute Interactions: An Approach Based on Entropy. Available at http://arxiv.org/abs/cs.AI/0308002 (Retrieved on November 27, 2008).
Knorr-Cetina, K. D. (1999). *Epistemic Cultures: How the Sciences Make Knowledge*. Cambridge, MA: Harvard University Press.
Kuhn, T. S. (1962). *The Structure of Scientific Revolutions*. Chicago: University of Chicago Press.
Latour, B. (1987). *Science in Action*. Milton Keynes: Open University Press.
Latour, B. & Woolgar, S. (1979). *Laboratory Life: The Social Construction of Scientific Facts*. London/Beverly Hills: Sage.
Leydesdorff, L. (1993). 'Structure'/'Action' Contingencies and the Model of Parallel Processing. *Journal for the Theory of Social Behaviour,* 23(1), 47-77.

Leydesdorff, L. (1995). *The Challenge of Scientometrics: The development, measurement, and self-organization of scientific communications.* Leiden: DSWO Press, Leiden University,
Leydesdorff, L. (1997). Sustainable Technological Developments and Second-Order Cybernetics. *Technology Analysis & Strategic Management,* 9(3), 329-341.
Leydesdorff, L. (1998). Theories of Citation? *Scientometrics,* 43(1), 5-25.
Leydesdorff, L. (2000). Luhmann, Habermas, and the Theory of Communication. *Systems Research and Behavioral Science,* 17(3), 273-288.
Leydesdorff, L. (2007). Scientific Communication and Cognitive Codification: Social Systems Theory and the Sociology of Scientific Knowledge. *European Journal of Social Theory* 10(3), 375-388.
Leydesdorff, L. (2008). Configurational Information as Potentially Negative Entropy: The Triple Helix Model. *Entropy,* 10(4), 391-410; available at http://www.mdpi.com/1099-4300/10/4/391 . (Retrieved on November 27, 2008.)
Leydesdorff, L., & Amsterdamska, O. (1990). Dimensions of Citation Analysis. *Science, Technology & Human Values,* 15, 305-335.
Leydesdorff, L., & Sun, Y. (2009). National and International Dimensions of the Triple Helix in Japan: University-Industry-Government versus International Co-Authorship Relations. *Journal of the American Society for Information Science and Technology* (in print).
Luhmann, N. (1986). The *Autopoiesis* of Social Systems. In F. Geyer & J. v. d. Zouwen (Eds.), *Sociocybernetic Paradoxes* (pp. 172-192). London: Sage
Luhmann, N. (1990). The Cognitive Program of Constructivism and a Reality that Remains Unknown. In W. Krohn, G. Küppers & H. Nowotny (Eds.), *Selforganization. Portrait of a Scientific Revolution* (pp. 64-85). Dordrecht: Reidel.
Luhmann, N. (1995). *Social Systems*. Stanford, CA: Stanford University Press.
Luhmann, N. (2002a). How Can the Mind Participate in Communication? In W. Rasch (Ed.), *Theories of Distinction: Redescribing the Descriptions of Modernity* (pp. 169–184). Stanford, CA: Stanford University Press.
Luhmann, N. (2002b). The Modern Sciences and Phenomenology. In W. Rasch (Ed.), *Theories of Distinction: Redescribing the descriptions of modernity* (pp. 33-60). Stanford, CA: Stanford University Press.
Maturana, H. R. (2000). The Nature of the Laws of Nature. *Systems Research and Behavioural Science,* 17, 459-468.
Maturana, H. R., & Varela, F. J. (1980). *Autopoiesis and Cognition: The Realization of the Living*. Boston: Reidel.
Maturana, H. R, & Varela, F. J. (1992). *The tree of knowledge: The biological roots of human understanding* (rev. ed.). Boston: Shambhala.
McGill, W. J. (1954). Multivariate information transmission. *Psychometrika,* 19(2), 97-116.
McGill, W. J., & Quastler, H. (1955). Standardized nomenclature: An attempt. In H. Quastler (Ed.), *Information Theory in Psychology: Problems and Methods* (pp. 83–92). Woodbury, NY: The Free Press.
Merton, R. K. (1968). *Social Theory and Social Structure*. New York.
Merton, R. K. (1973). *The Sociology of Science: Theoretical and Empirical Investigations.* Chicago: University of Chicago Press.
Merton, R.K. (1979). "Foreword," in: E. Garfield, *Citation Indexing – Its Theory and Application in Science, Technology, and Humanities*. Philadelphia: ISI Press, pages v-vi.

Moed, H. F., Glänzel, W., & Schmoch, U. (2004). *Handbook of Quantitative Science and Technology Research: The Use of Publication and Patent Statistics in Studies of S & T Systems*. Dordrecht, etc.: Springer.

Mulkay, M., Potter, J., & Yearley, S. (1983). Why an Analysis of Scientific Discourse is Needed. In K. D. Knorr & M. J. Mulkay (Eds.), *Science Observed: Perspectives on the Social Study of Science* (pp. 171-204.). London: Sage.

Myers, G. (1985). Texts as Knowledge Claims: The Social Construction of Two Biology Articles. *Social Studies of Science,* 15, 593-630.

Nelson, R. R., & Winter, S. G. (1977). In Search of Useful Theory of Innovation. *Research Policy,* 6, 35-76.

Nelson, R. R., & Winter, S. G. (1982). *An Evolutionary Theory of Economic Change*. Cambridge, MA: Belknap Press of Harvard University Press.

Pinch, T. J. (1985). Towards an Analysis of Scientific Observation: The Externality and Evidential Significance of Observational Reports in Physics,. *Social Studies of Science* 15, 3-36.

Popper, K. R. ([1935] 1959). *The Logic of Scientific Discovery*. London: Hutchinson.

Price, D.J. de Solla (1965). Networks of scientific papers. *Science* 149 (3683), 510-515

Price, D. J. de Solla (1970). Citation Measures of Hard Science, Soft Science, Technology, and Nonscience. In C. E. Nelson & D. K. Pollock (Eds.), *Communication among Scientists and Engineers* (pp. 3-22). Lexington, MA: Heath.

Simon, H. A. (1973). The Organization of Complex Systems. In H. H. Pattee (Ed.), *Hierarchy Theory: The Challenge of Complex Systems* (pp. 1-27). New York: George Braziller Inc.

Small, H. (1978). Cited Documents as Concept Symbols. *Social Studies of Science*, 8(3), 327-340.

Smolensky, P. (1986). Information Processing in Dynamical Systems: Foundation of Harmony Theory. In D. E. Rumelhart, J. L. McClelland & the PDP Research Group (Eds.), *Parallel Distributed Processing* (Vol. I, pp. 194-281). Cambridge, MA/ London: MIT Press.

Spiegel-Rösing, I. (1973). *Wissenschaftsentwicklung und Wissensschaftssteuerung*. Frankfurt a.M.: Athenaeum Verlag.

Stichweh, R. (1990). Self-Organization and Autopoiesis in the Development of Modern Science. In: Krohn, W., Küppers, G., Nowotny, H. (Eds.), *Self-Organization: Portrait of a Scientific Revolution*. Dordrecht: Reidel, pp.195-207.

Sun, Y., & Negishi, M. (2008). Measuring relationships among university, industry and the other sectors in Japan's national innovation system. *10th International Conference on Science and Technology Indicators* , J. Gorraiz & E. Schiebel (Eds.), Vienna, 17-20 September 2008: Austrian Research Centers, pp. 169-171.

Theil, H. (1972). *Statistical Decomposition Analysis*. Amsterdam: North-Holland.

Ulanowicz, R. E. (1997). *Ecology, The Ascendent Perspective*. New York: Columbia University Press.

White, H.D., Wellman, B. & Nazer, N. (2004). Does citation reflect social structure? *Journal of the American Society for Information Science*, 55(2), 111-26.

Whitley, R. (1984). *The Intellectual and Social Organization of the Sciences*. Oxford: Clarendon Press.

Wouters, P. (1999). *Beyond the Holy Grail: From citation theory to indicator theories.* Scientometrics 44 (3) : 561-580.

Yeung, R. W. (2008). *Information Theory and Network Coding.* New York, NY: Springer; available at http://iest2.ie.cuhk.edu.hk/~whyeung/post/main2.pdf (Retrieved on November 11, 2008).